\documentclass{PoS}

\newcommand{\ms}{M$_{\odot}$}
\newcommand{\zs}{Z$_{\odot}$}

\title{On the early chemical evolution of the Milky Way}

\ShortTitle{Milky Way evolution}

\author{\speaker{Nikos Prantzos}\\
        Institut d'Astrophysique de Paris\\
        E-mail: \email{prantzos@iap.fr}}

\abstract{A few topics concerning the early chemical evolution of the Milky Way are critically discussed. In particular, it is argued
that: 1) Observed abundance patterns of extremely metal poor stars (of Pop. II) do not constrain the mass range of the first generation (Pop. III) stars;
the latter may well be normal massive stars (10-50 \ms) or very massive ones (140-1000 \ms) or a combination of the two classes. 2) The discrepancy between primordial Li abundance (after WMAP) and the observed ``Spite plateau''  cannot be  due to astration by a generation of massive Pop. III stars, as recently suggested, unless if such stars eject negligible amounts of metals. 3) The observed halo metallicity disribution may well be understood in the framework of hierarchical galaxy formation, as shown here with a simple semi-analytical model. 4) Formation of the Milky Way's halo from a myriad of smaller sub-haloes may have important implications for our understanding of the abundance patterns  of r-elements, the origin of which remains still unclear.}

\FullConference{International Symposium on Nuclear Astrophysics --- Nuclei in the Cosmos --- IX\\
		 June 25-30 2006\\
		 CERN, Geneva, Switzerland}
\usepackage{graphicx}
\begin{document}

\section{Introduction}

Galactic chemical evolution studies provide  a useful
framework for the interpretation of the everincreasing wealth
of observational data concerning abundances in stars of the Milky Way.
In particular, studies of abundance ratios in the oldest stars of the 
Milky Way (Pop. II stars, with metallicities down to [Fe/H]=log(Fe/H)-log(Fe/H)$_{\odot}$=--4.) 
allow one to put interesting constraints on the nature of the first
stars that enriched the interstellar medium with metals. Ultimately, 
one may hope to find a distinctive imprint of the very first  
generation of massive stars (Pop. III), those that were born out of
primordial material containing H and He only (as well as trace amounts 
of $^7$Li, resulting from the Big Bang).  This pattern should then be visible on the
surfaces of the next generation stars that were born from the enriched
gas and are still around today, i.e.
the oldest and most metal poor Pop. II stars.

\section{Halo star composition: any signatures of massive Pop. III stars ?}
 
Star formation theory and numerical simulations suggest that
the first stars were very massive (e.g. Nakamura and Umemura 2002,
Glover 2004 and references therein).
 In view of the many important implications
of those ideas (e.g. for re-ionization of the Universe)  it is interesting to investigate
whether the observed abundance 
patterns of old stars reveal any trace of a  Pop III stellar generation with
peculiar properties. In fact, it has been suggested that a stellar initial mass function (IMF) rich in
massive stars (top-heavy IMF) can help explain (1) the abundance ratios in the Fe-peak region, and (2) the discrepancy between the expected primordial Li value (after the determination of the cosmic baryonic density by WMAP) and the one observed in the ``Spite plateau''. We argue below that the answers are: (1) may be, and (2) no.

\subsection{Abundance patterns in EMP stars: no clear signature of the culprits !}
\begin{figure}
\centering
\includegraphics[height=\textwidth,angle=-90]{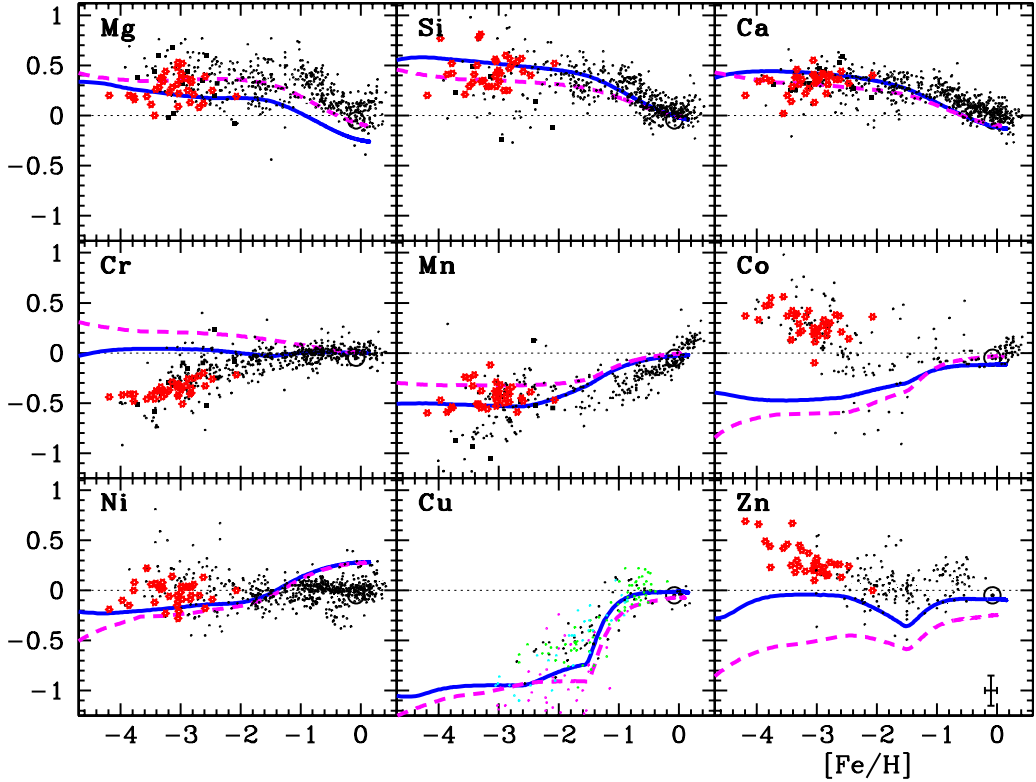}
\caption{
Abundance ratios [X/Fe] as a function
of metallicity [Fe/H] in stars of the Milky Way; small data points
are from various sources, while the large data points at low 
metallicity are from the  VLT survey of Cayrel et al. (2004).
Results of standard chemical evolution models, performed with two
sets of metallicity dependent massive star yields 
(Woosley and Weaver 1995, {\it solid curves} ; 
Chieffi and Limongi 2004, {\it dashed
curves}) are also displayed. Yields for SNIa are adopted from 
Iwamoto et al. (1999) and for intermediate mass stars from 
van den Hoek and Gronewegen  (1997). While the behaviour of alpha elements
and Mn is correctly reproduced (at least qualitatively), 
there are large discrepancies
in the cases of Cr, Co and Zn.}
\label{fig:1}       
\end{figure}

The recent data on Extremely metal Poor (EMP) stars
confirm some previously found tendencies in the abundance patterns of heavy elements, and in particular:

- The constancy, down to the lowest observed metallicities,
of the $\alpha$/Fe ratio (where $\alpha$ stands for alpha elements like
O, Mg, Si, Ca); the high value of that ratio ($\sim$3 times solar) is
attributed to the late contribution of Fe by SNIa, with long lived progenitors.

- The decreasing trend of Mn/Fe and Cu/Fe with decreasing metallicity, in
qualititative agreement with theoretical expectations (see Fig. 1).

However, more interesting are the cases where observations are at odds with
theoretical expectations. These are the cases of 
Cr, Co and Zn, as seen in
Fig. 1, where the results of
a full-scale chemical evolution model of the Milky Way (originally presented
in Goswami and Prantzos 2000) are displayed. The adopted stellar yields
are from non-rotating Core Collapse supernovae (CCSN) in the ``normal'' mass range
of 12 to $\sim$40 \ms, exploding with ``canonical'' energies of $E=$10$^{51}$erg and with progenitor metallicities from Z=0 to
Z=\zs.

The reasons for those discrepancies probably lie in our poor understanding of
the explosion mechanism and of the nature of the early CCSN.
A large amount of work was devoted in recent years to understanding the peculiar abundance patterns among Fe-peak elements observed in EMP stars (e.g. Heger and Woosley 2002, Maeda and Nomoto 2003, Chieffi and Limongi 2004; for reviews see Nomoto et al. 2005,  Cayrel 2006 and references therein). In those works it was explored whether some specific 
property of the Pop. III stars (characteristic mass, rotation, explosion energy 
etc.) was quite different from the corresponding one of their more metal
rich counterparts and produced a different nucleosynthetic
pattern in the ejecta.
Two main classes of solutions (not mutually exclusive) appear to arise from those works:

\begin{figure}[t!]
 \includegraphics[width=4.5cm,angle=-90]{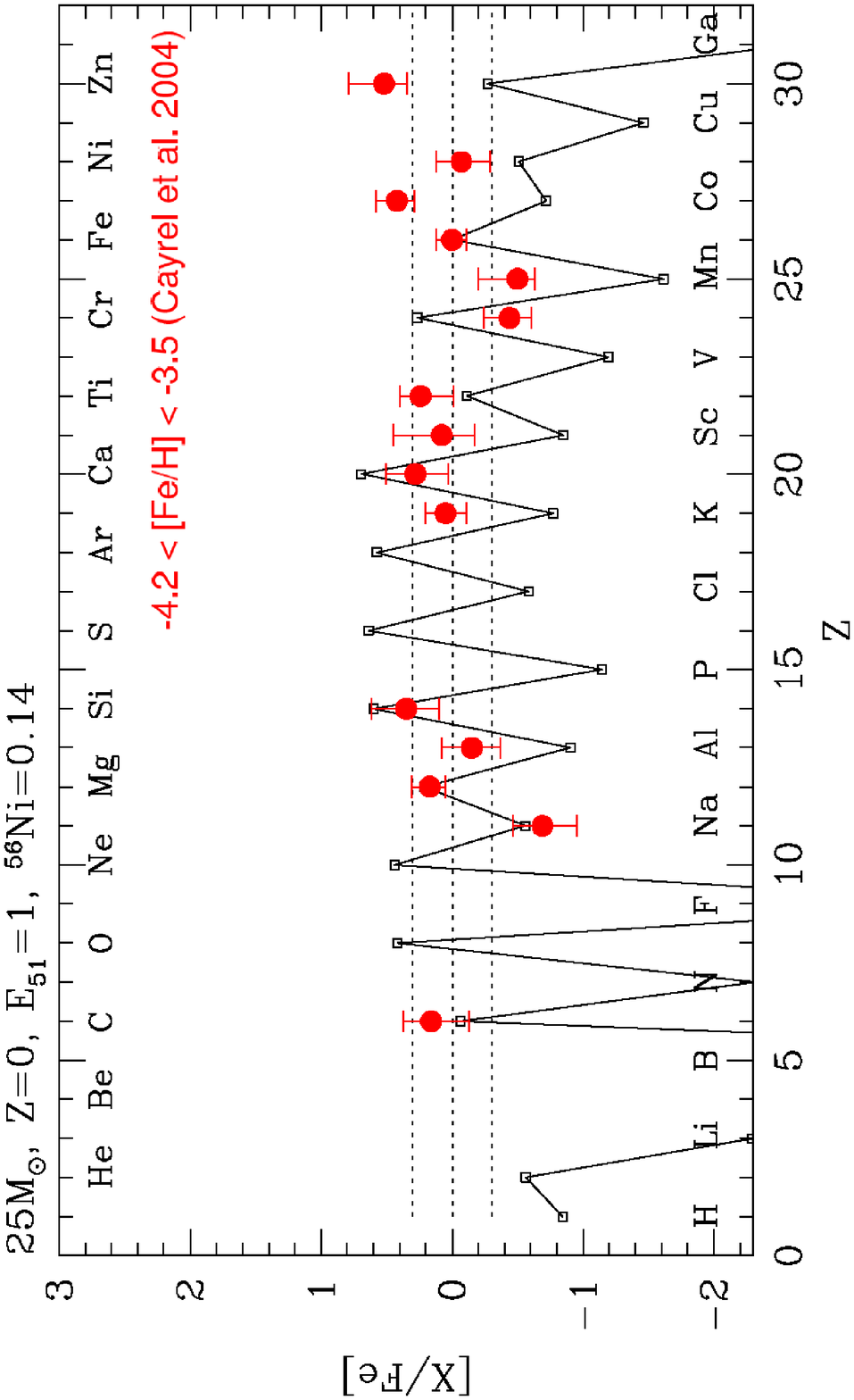}
\qquad
 \includegraphics[width=4.3cm,angle=-90]{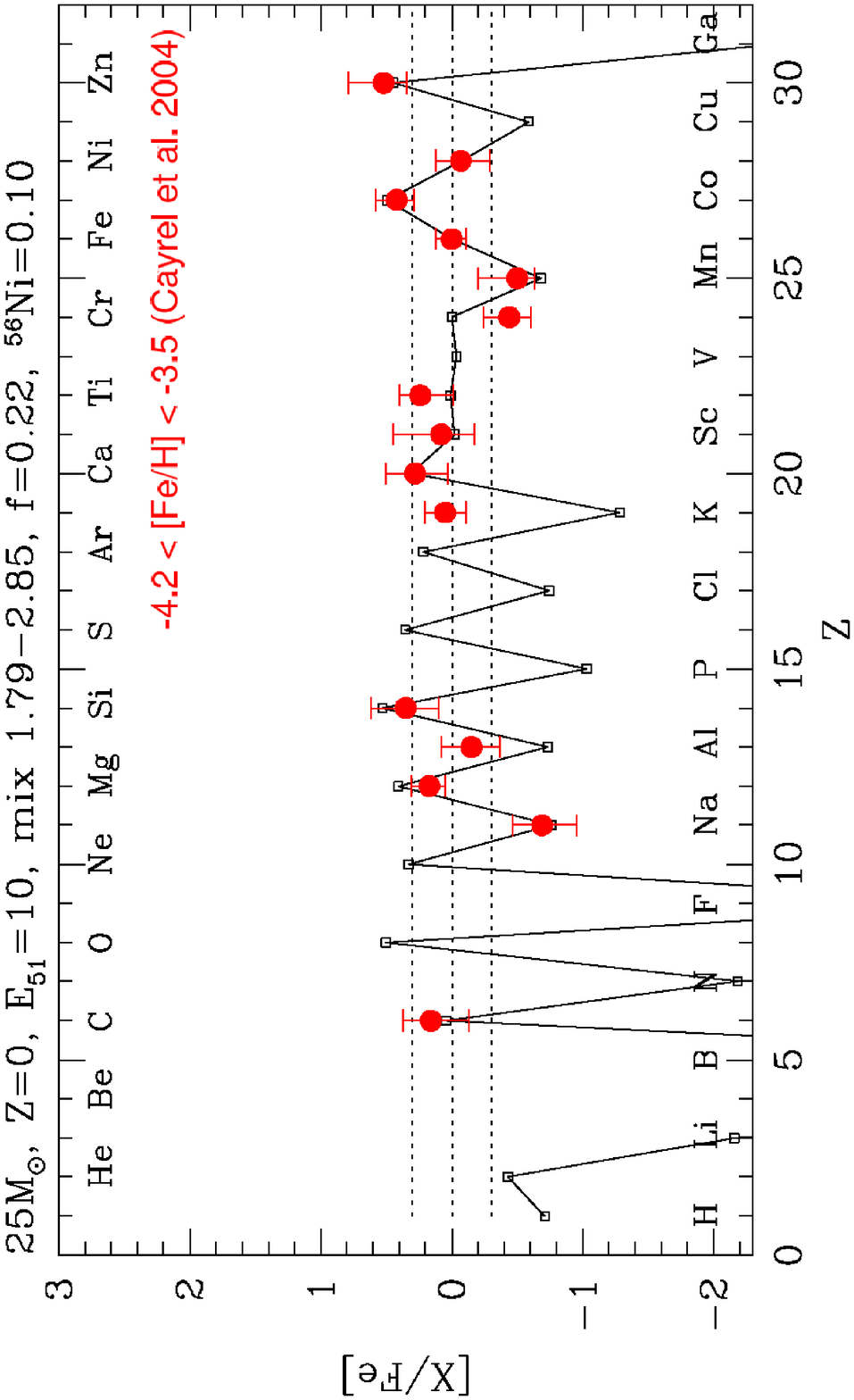}
\caption{Nucleosynthesis yields of a 25 \ms \ star, compared to observations of extremely metal-poor stars (vertical error bars, from Cayrel et al. 2004). {\it Left:} The star explodes with a canonical energy of 1.5 10$^{51}$ erg; {\it Right}: The star explodes as a hypernova, with an energy of 10 10$^{51}$ erg (both figures from Nomoto et al. 2005).   Abundance ratios in the Fe-peak are much better reproduced in the second case.
}
\label{fig:1}       
\end{figure}

1) {\it Normal massive stars (10-50 \ms)}, exploding as core collapse SN (CCSN) and leaving behind a neutron star or a black hole. It was shown (essentially by the Tokyo group, see Nomoto et al. 2005 for a recent review) that {\it higher energies} than the canonical one of $E=10^{51}$ ergs  {\it combined with asphericity}
of the explosion can help to improve the situation concerning the Zn/Fe and Co/Fe ratios observed in EMPs. Material ejected along the rotation  axis (in the form of jets) has high entropy
and is found to be enriched in products of a-rich freeze-out (Zn and Co), as well as Sc (which is 
generically undeproduced in spherical models); this kind of models seems at present
the most promising, but this is not quite unexpected (since they have 
at least one more degree of freedom w.r.t. spherical models). Such properties (high energies, asphericity) are indeed observed in the local Universe for {\it hypernovae}, a class of energetic CCSN. However, hypernovae are not so frequent today as to affect the late chemical evolution of galaxies (normal CCSN plus SNIa can easily account for e.g. the solar abundance pattern). If hypernovae indeed affected the early chemical evolution, they muct have been much more abundant in  Pop. III than today. No satisfactory explanation for such a large hypernova fraction in the early Universe exists up to now. It seems hard to conceive that the typical energy of the explosion was substantially higher in Pop. III stars (all other parameters, except metallicity, being kept the same). On the other hand, rotation effects may indeed be stronger in low metallicity environments, because of smaller angular momentum losses due to lower mass loss rates (e.g. Meynet et al. 2006). In any case, none of the models explored sofar appears to account for the early trend of Cr/Fe, which remains a  mystery at present.

2) {\it Very massive stars (VMS)}, above 100 \ms. Such stars are thought to: collapse to black holes, if M$<$140 \ms; explode during oxygen burning as pair-implosion SN (PISN), if 140$<$M/\ms$<$300; and again collapse without ejecting metals if M$>$300 \ms \ (see Heger and Woosley 2002 and references therein). In the case of PISN, it was shown that they do not produce enough Zn to account for the oberved high Zn/Fe ratio in EMP stars (Heger and Woosley 2002, Umeda and Nomoto 2005) and they were thus excluded as major contributors to the early chemical evolution of the MW. This shortcoming cast doubt as to the existence of a Pop. III generation composed exclusively of VMS. However, in a recent work Ohkubo et al. (2006) explore the conditions under which {\it rotating stars} of M$=$500  and 1000 \ms \ {\it do explode and eject metals} (although a massive black hole is ultimately produced in both cases). They find explosions for a certain region of their parameter space and substantial production of Zn as a generic feature (see Fig. 3b). Even if each class of PISN or 500-1000 \ms \ stars cannot reproduce the observed EMP abundance pattern alone, it is obvious that considering the full 140-1000 \ms \ range (folded with an appropriate IMF) will account for such  a pattern. Thus, {\it VMS cannot be excluded at present as candidates for Pop. III stars}, at least not on ``chemical'' grounds.

\begin{figure}
 \includegraphics[width=4.8cm,angle=-90]{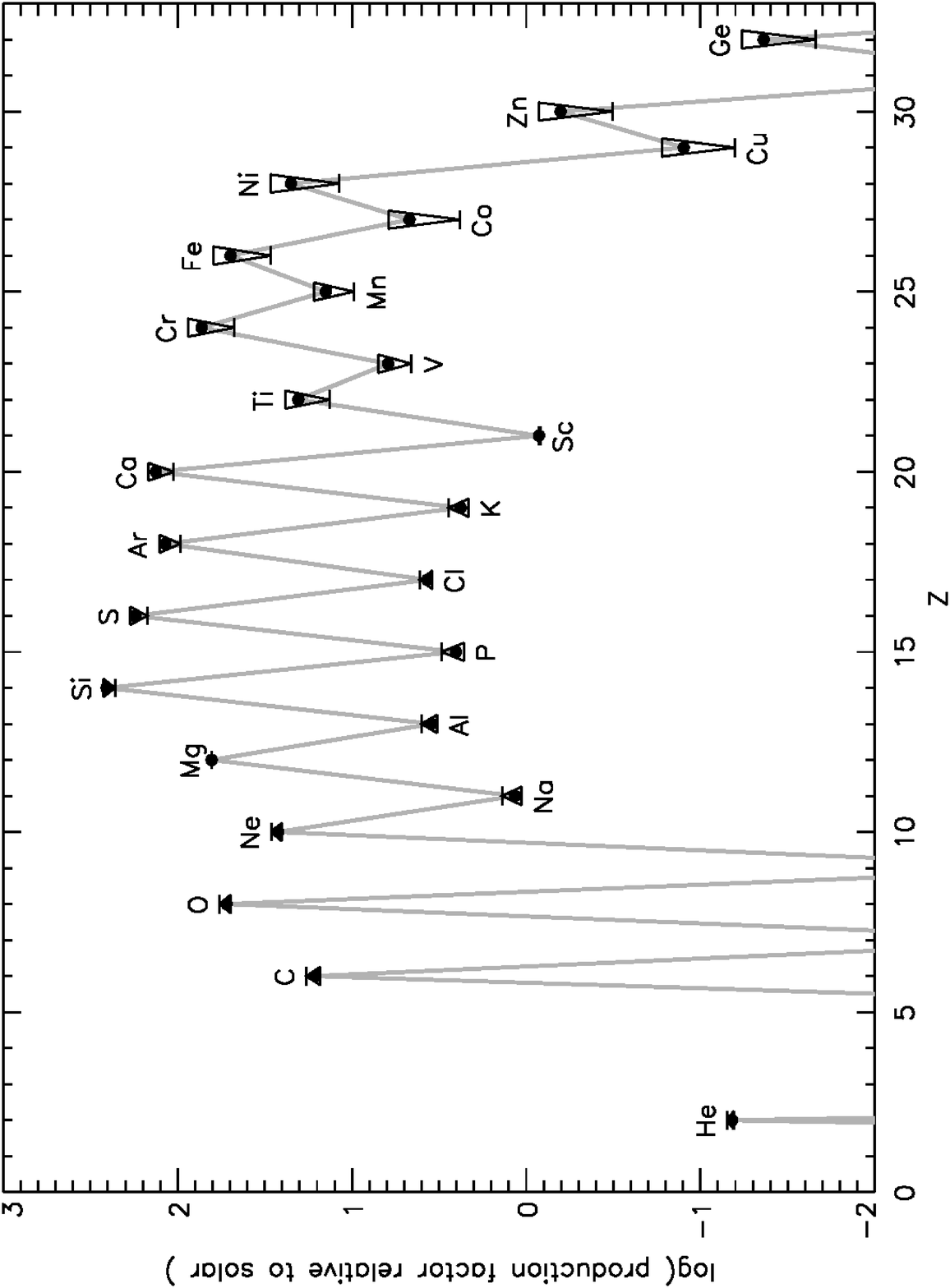}
\qquad
 \includegraphics[width=4.9cm,height=7.8cm,angle=-90]{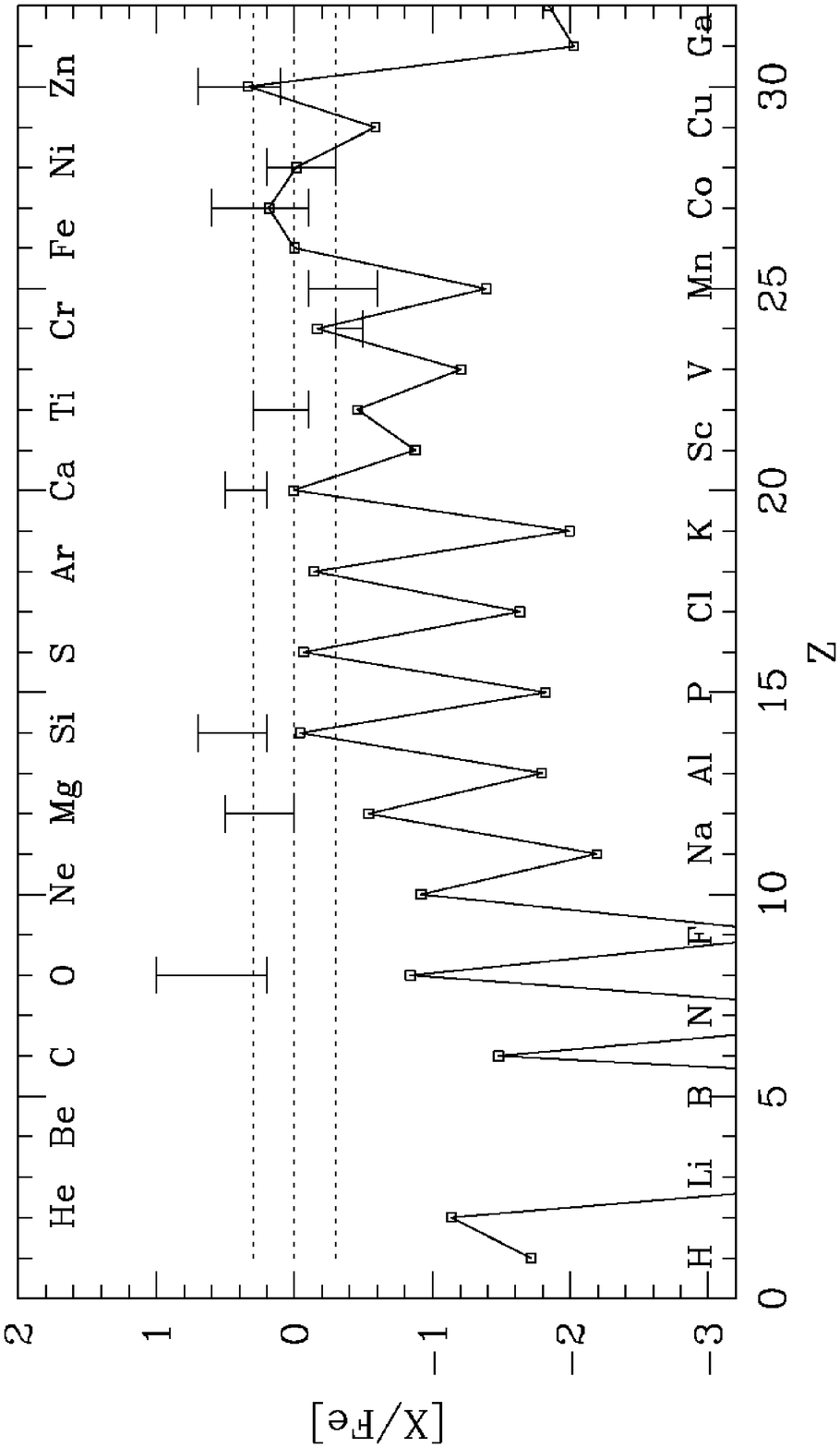}
\caption{Nucleosynthesis yields of Very Massive Stars (VMS, with mass $>$100 \ms). {\it Left:}  Stars in the 140 to 260 \ms \ range  explode as Pair Instability Supernovae (PISN) and do not produce enough Zn (from Heger and Woosley (2002). {\it Right}: Stars in the 300-1000 \ms \ range explode after core collapse and produce significant amounts of Zn (from Ohkubo  et al. 2006).   Abundance ratios in the Fe-peak are much better reproduced in the second case.
A Pop. III population composed exclusively of VMS (in the 140-1000 \ms \ range) and with an appropriate IMF is obviously compatible with observations of EMP stars.
}
\label{fig:1}       
\end{figure}

\subsection {WMAP Li was (most probably) NOT astrated in  massive Pop. III stars}

The recent WMAP results from the 3-year data analysis of the cosmic microwave background, combined with other cosmological measurements, allow to determine with great precision the parameters characterising our observable Universe (see Lahav and Liddle 2006). In particular, the baryonic density is $\Omega_B$=4.5 10$^{-2}$. For that density, calculations of Standard Big Bang Nucleosynthesis (SBBN) predict an amount of primordial deuterium that is fully compatible with observations of its abundance in remote gas clouds (at such early times that significant astration had no time to occur); this undoubtely constitutes a triumph for SBBN (e.g. A. Coc, this Conference). Unfortunately, the situation with Li ($^7$Li) is far less satisfactory.

\begin{figure}
\includegraphics[height=0.45\textwidth,angle=-90]{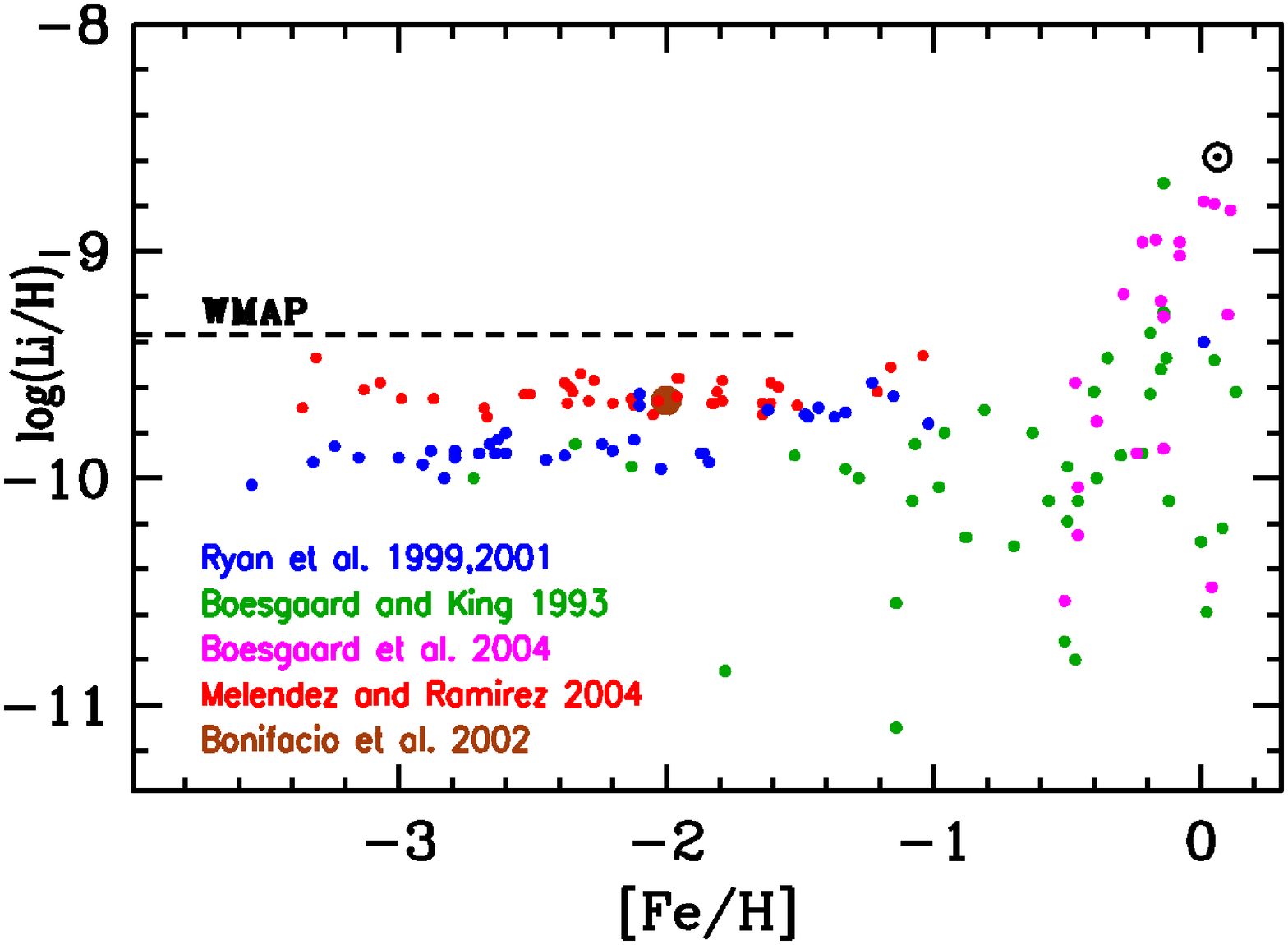}
\qquad
\includegraphics[height=0.45\textwidth,angle=-90]{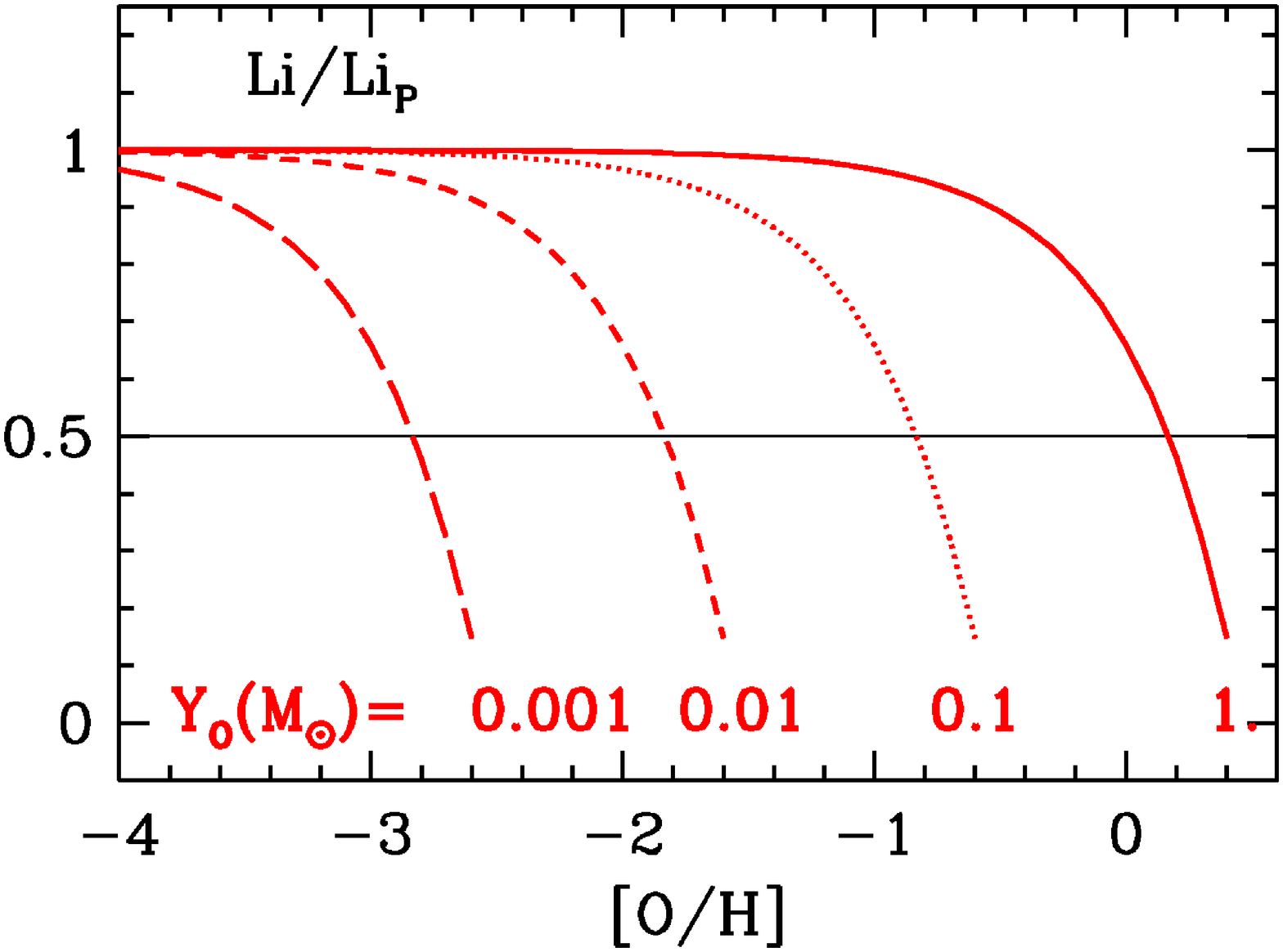}
\caption{{\it Left:} Observations of Li in halo and disk stars of the Milky Way. The primordial Li value, obtained from the baryonic density of WMAP and calculations of SBBN, is indicated by a {\it horizontal dashed line}. The observed Li plateau at low metallicities depends sensitively on assumed stellar temperature, and differs from the WMAP value by factors 2-3. {\it Right:} Illustration of the problem encountered by the idea that the discrepancy is due to astration of Li by a Pop. III composed of exclusively massive stars, in the 10-100 \ms \ range (Piau et al. 2006): such stars necessarily eject metals, either through their winds (e.g. Nitrogen, in case of rotating stars) or through the final supernova explosion (e.g. Oxygen). Mixing the ejecta with various proportions of primordial material would result in Li depletion (by a factor of $\sim$2 in case of a 50-50 mixture), but the resulting metallicity would be much higher than the one of EMP stars unless abnormally low oxygen yields were assumed (i.e. the curve in Fig. 4b parametrized with a yield $Y_O$=0.001 \ms, see text). 
}
\label{fig:4}       
\end{figure}

Indeed, the Li abundance of the "Spite plateau" (Li/H$\sim$ 1.5 10$^{-10}$=const. for halo stars, down to the lowest metallicities) is a factor of 2-3 lower than the WMAP+SBBN value. Barring systematic errors, the conclusion is that primordial Li has been depleted, either {\it before} getting into the stars it is observed today, or {\it during} the lifetime of those stars. Two "depletion agents" have been proposed in the former case: decaying supersymmetric particles (Jedamzik 2006) and astration in a first generation of exclusively massive (mass range $m_*$=10-40 \ms) Pop. III stars (Piau et al. 2006). The latter idea,
however, suffers from a serious flaw, since in that case the metallicity of the ISM (out of which  the next stellar generation would form with depleted Li) would necessarily rise to levels much higher than those observed in EMP stars. This can be seen as follows:

Assuming that the current halo stellar mass ($M_H$=2 10$^9$ \ms) was initially in the form of gas, a fraction $f$ of which was astrated through massive stars,  the resulting Li mass fraction is $X_{Li}$=$(1-f) X_{Li,P}$ where $X_{Li,P}$ is the primordial Li abundance (assuming a return fraction $R\sim$1 for the astrating and metal producing stars). Similarly, the resulting oxygen abundance would be $X_O$=$m_O$/$M_H$, where the mass of oxygen $m_O=N_{SN} \ Y_O$ is produced by a number of supernovae $N_{SN}$= $f M_H/ m_*$, each one with a typical oxygen yield of $Y_O$ (in \ms, to be discussed below). Putting  all this together, one has: 
$${{X_{Li}}\over{X_{Li,P}}} \ = \ 1 - 0.28 {{(X_O/0.007) (m_*/40 M_{\odot})}\over{Y_O}} $$
That relation appears in Fig. 4 (right panel) as a function of log($X_O/X_{O,\odot}$), with solar $X_{O,\odot}$=0.007. The four  curves correspond to different assumptions about the typical oxygen yield of a massive star of Z=0, ranging from 0.001 to 1 \ms. Among the various calculations
of massive star nucleosynthesis made sofar, only one set of models in the 30-40 \ms \ range (set A of Woosley and Weaver 1995, WW95) produces extremely low oxygen (and heavy element) yields for stars of 35 and 40 \ms. This is due to the assumption of constant kinetic energy at infinity (=1.2 10$^{51}$ erg) adopted for all models of set A in WW95; other assumptions (sets B and C) lead to much larger oxygen yields and to abundance ratios in rough agreement with observations in EMP stars, at least for alpha-elements (e.g. Goswami and Prantzos 2000), while set A provides a rather poor match. Other calculations at Z=0, either for spherically symmetric stars (Chieffi and Limongi 2004, assuming a fixed mass of $^{56}$Ni ejected) or for bipolar explosions (Maeda and Nomoto 2003) lead to $Y_O>$1 \ms \ always; note that the latter models  provide a better (albeit not perfect) match to observed Fe-peak element abundance ratios than ``standard'' (spherically symmetric) models, as dicussed in Sec. 2.1.

Another way to eject astrated material by Z$\sim$0 stars is through stellar winds, which require rapidly rotating stars, 
since radiative pressure is inneficient at such low  metallicities;  but rotating massive stars produce large amounts of nitrogen (which may in fact help explaining the observed primary-like nitrogen in EMP stars, e.g. Meynet et al. 2006), thus the problem of metal overproduction is not avoided in that case either.

Astration in massive Pop. III stars cannot solve the Li discrepancy between the Spite plateau and WMAP+SBBN \footnote{At least, not the 10-40 \ms stars suggested in Piau et al. 2006 (provided that current nucleosynthesis models for such stars are correct); 100 \ms \ stars collapsing to black holes would be better candidates (provided they eject a substantial fraction of their astrated mass).}: even a small Li depletion should be accompanied by excessive  metal enhancement. It appears now that  depletion {\it during} the stellar evolution, within the stellar envelope, is the answer (see K\"orn et al. 2006 and references therein).

\section{The early MW and  hierarchical galaxy formation}

In the previous sections, the early chemical evolution  of the MW was discussed independently of the cosmological framework in which it took place.
According to the currently dominant paradigm of hierarchical structure formation, the early phases of a galaxy's evolution are the most complex ones, as they involve multiple mergers of smaller sub-units. In the case of the Milky Way,  interesting "chemical signatures" of that period 
should still be left around us today, in the form of the metallicity distribution (MD) of long-lived stars and in dispersion in abundance ratios. We discuss those two topics in the following.

\begin{figure}
\includegraphics[height=6.5cm]{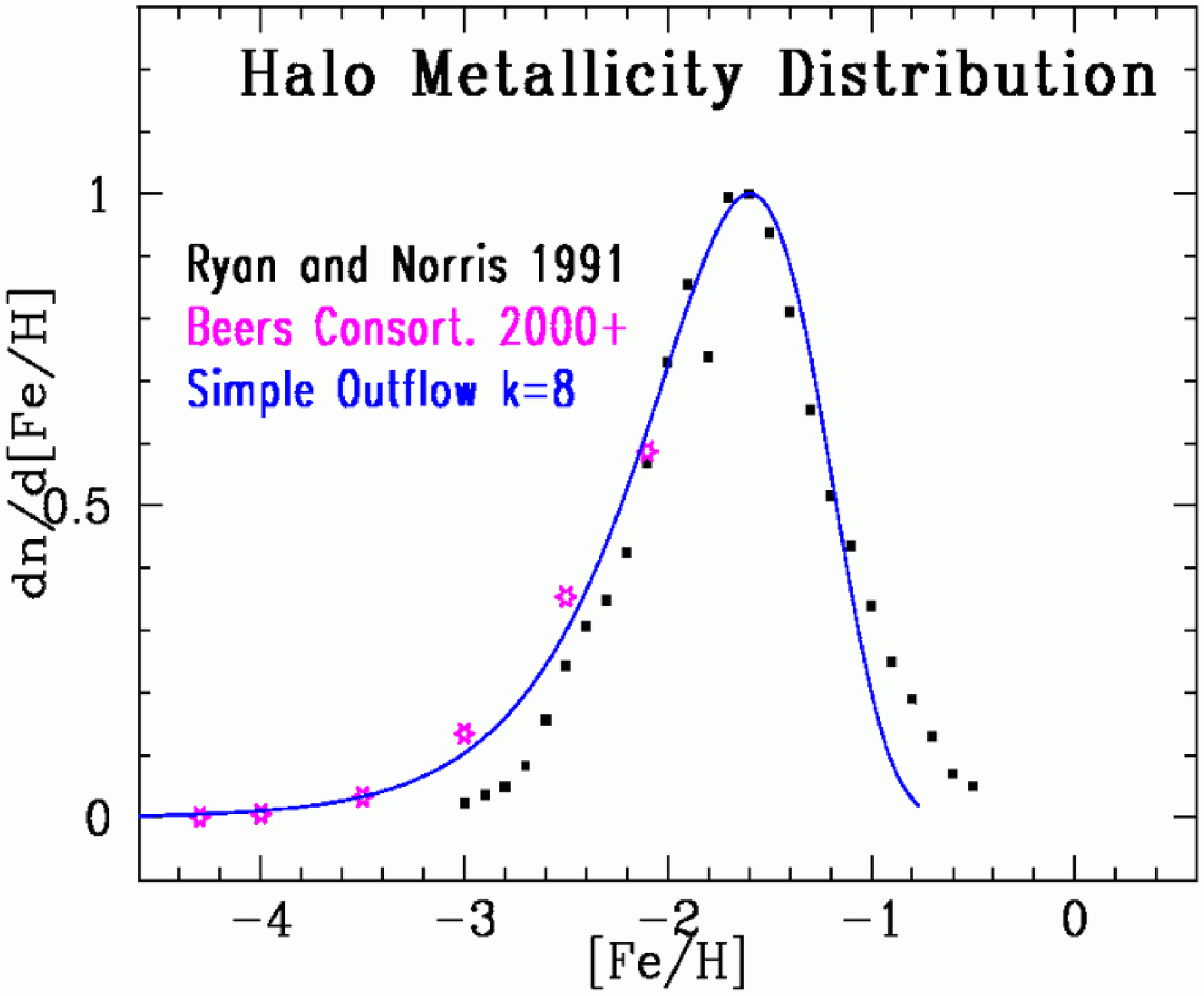}
\qquad
\includegraphics[height=6.5cm]{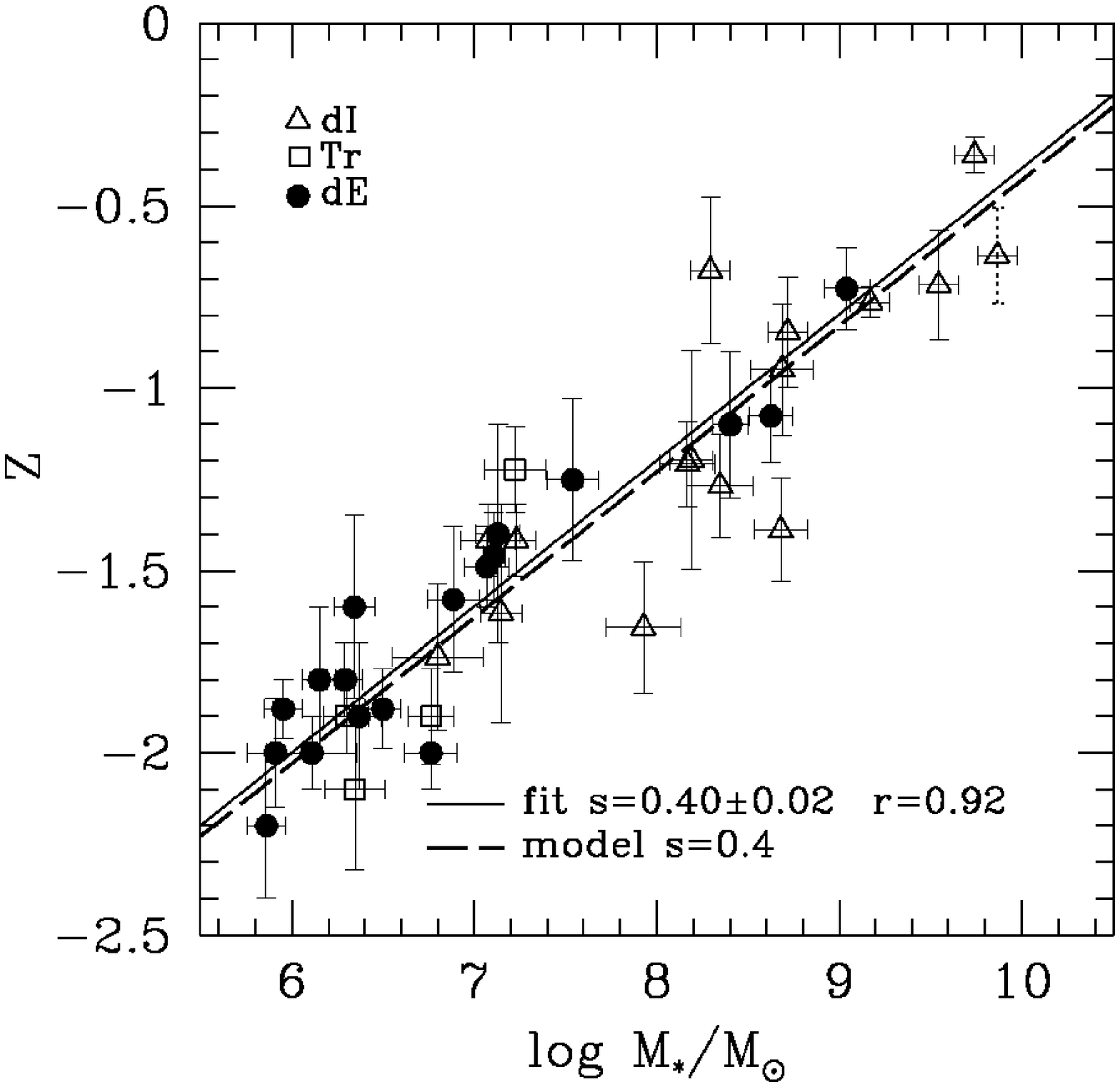}
\caption{{\it Left:} Metallicity distribution of field halo stars from Ryan and Norris (1991, dots), and the ongoing research of Beers and collaborators (T. Beers, private communication, asterisks). The curve is a simple model with outflow rate equal to 8 times the star formation rate. {\it Right:} Stellar metallicity vs stellar mass for neary galaxies. Data and model are from Dekel and Woo (2003). The MW halo, with average metallicity [Fe/H]$\sim$-1.6 (see left panel) and estimated mass 2 10$^9$ \ms \ falls below that relationship. }
\label{fig:3}       
\end{figure}

\subsection{The halo metallicity distribution: outflow vs. subhalo merging}

The MD of Galactic halo field stars (HMD) is rather well known in the metallicity range --2.2$<$[Fe/H]$<$--0.8, while at lower metallicities its precise form has still to be established by ongoing surveys (Fig. 5a). Its overall shape is well fitted  by a simple model of GCE as $dn/dlogZ \propto Z/y \ e^{-Z/y}$, where $y$ is the $yield$ of a stellar generation; this function has a maximum for $Z=y$. The HMD peaks at a metallicity [Fe/H]=--1.6 (or [O/H]=--1.1, assuming [O/Fe]$\sim$0.5 for halo stars), pointing to a low yield $y$=1/13 of the corresponding value for the solar neighborhood. Such a low halo yield is "classicaly" (i.e. in the monolithic collapse scenario) interpreted as  due to $outflow$ during halo formation (Hartwick 1976), at the large rate of 8 times the SFR (Prantzos 2003). How can it be understood in the framework of hierarchical merging ?

\begin{figure}
\includegraphics[width=6.9cm]{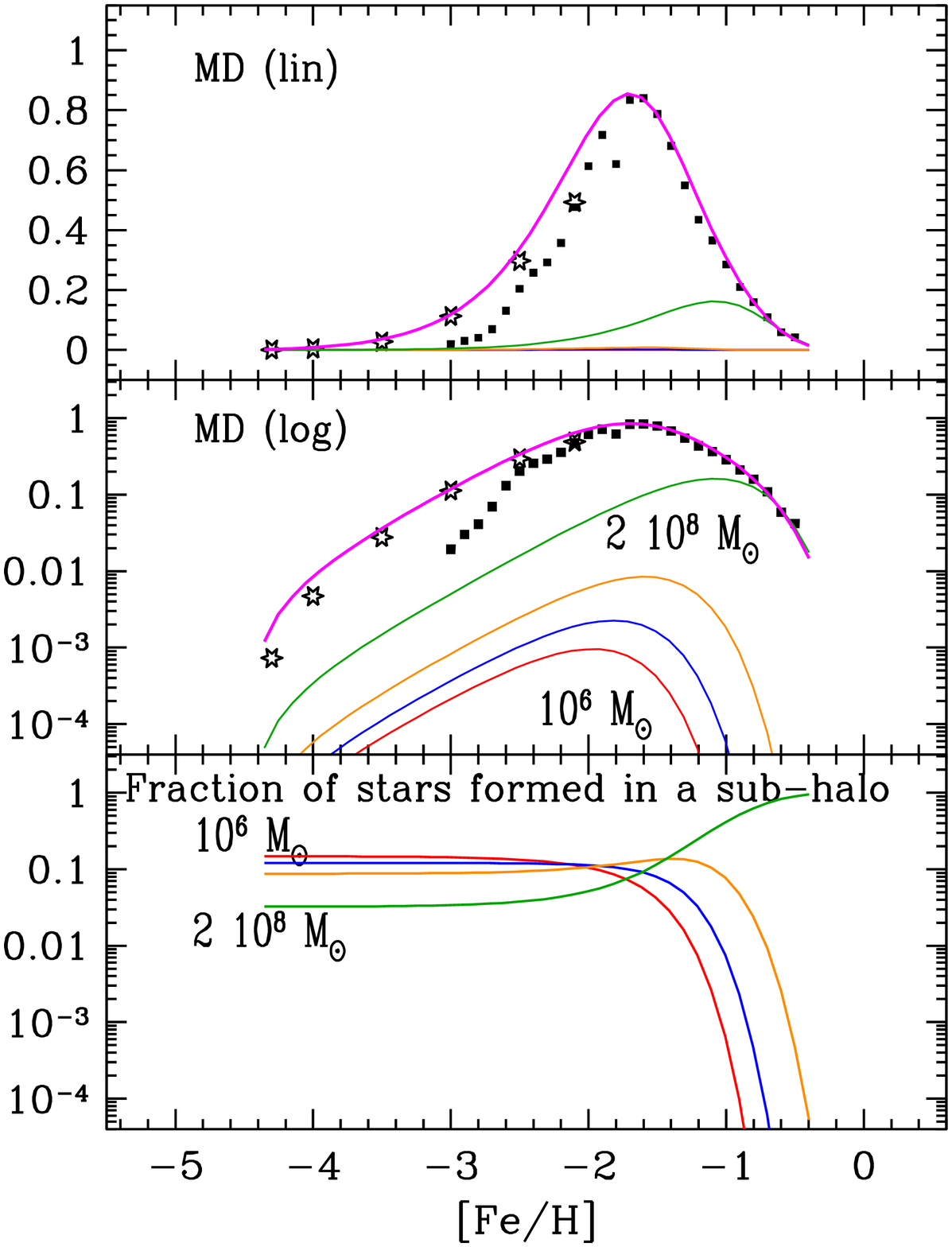}
\qquad
\includegraphics[width=6.9cm]{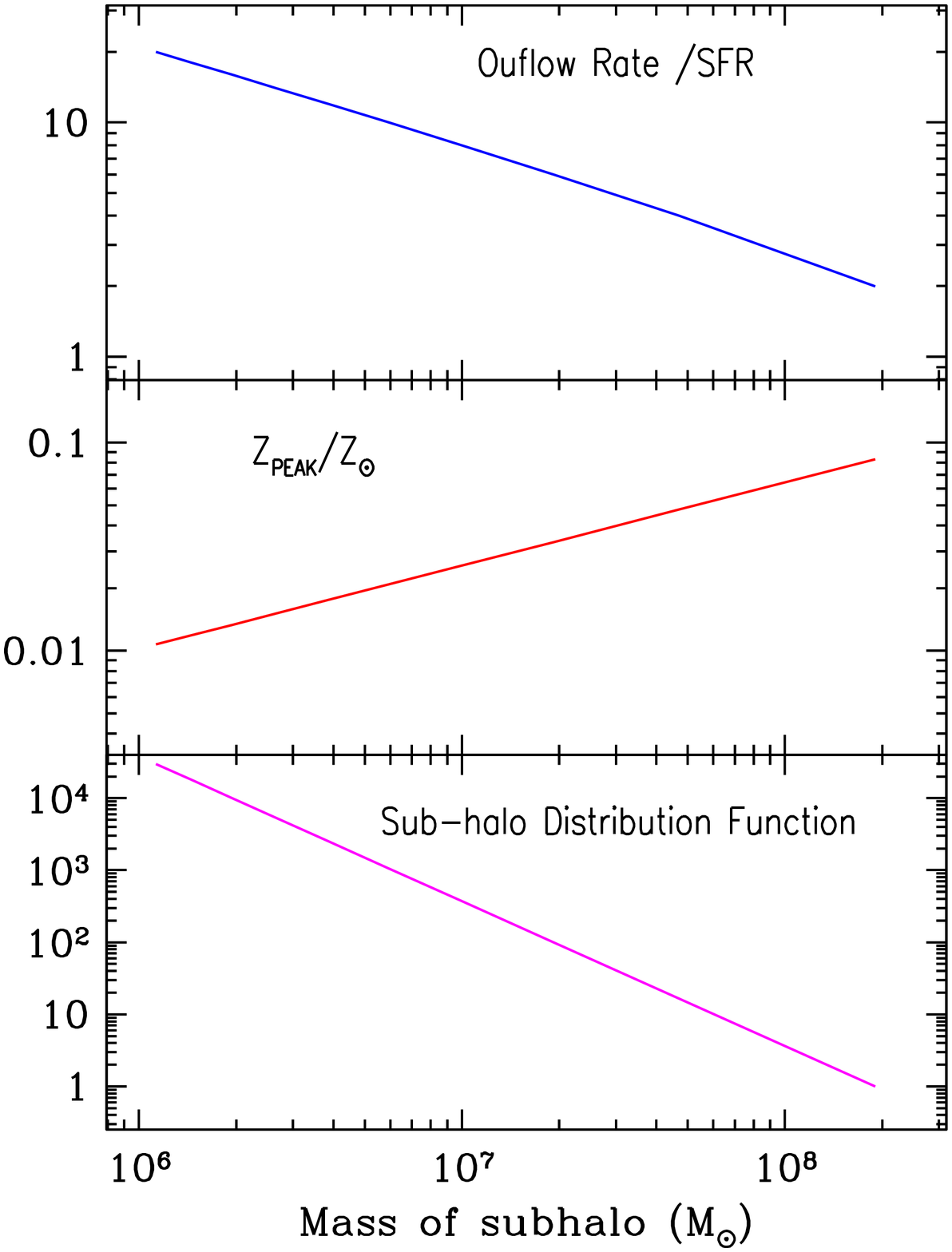}
\caption{{\it Left top} and {\it middle} panels: Metallicity distribution (in lin and log scales, respectively) of the MW halo, assumed to be composed of a population of smaller units (sub-haloes). The individual MDs of the sub-haloes, from 10$^6$ \ms to 2 10$^8$ \ms are indicated in both panels (but clearly seen only in the middle), as well as the sum over all haloes (upper curves in both panels, compared to observations). Small sub-haloes contribute the largest fraction of the lowest metallicity stars ({\it bottom left}).  Properties of the sub-haloes as a function of their mass.}
\label{fig:3}       
\end{figure}

It should be noted that the typical halo metallicity ([Fe/H]=-1.6) is substantially lower, by more than 0.5 dex,  than the corresponding metallicities of nearby  galaxies of similar mass ($M_{Halo}\sim$2 10$^9$ \ms), as can be seen in Fig. 5b. That figure also displays the well known galaxian relationship between stellar mass and stellar  metallicity; most probably, that relationship results from mass loss, which is more important in lower mass galaxies, since the hot supernova ejecta escape more easily their swallow potential well (e.g.  Dekel and Woo 2003). 

Assuming that the MW halo has been assembled from sub-units similar to the low mass galaxies of Fig. 5b, one may interpret the HMD as the sum of  the MD of such low mass galaxies; it is assumed that each one of those galaxies evolved with an appropriate outflow rate and corresponding effective yield $y(M)=(1-R)/(1+k-R)$, where $k(M)$ is the outflow rate in units of the SFR and $R$ is the return mass fraction,  depending on the adopted stellar IMF (see Prantzos 2003). In that case, one has: HMD($Z$) = 1/$M_{Halo}$ $\int Z/y(M) \ e^{-Z/y(M)} \Phi(M) M dM$, where $\Phi(M)$ is the mass function of the sub-units  and $y(M)$  is the effective yield of each sub-unit (obtained directly from Fig. 5b as $y(M)=Z(M)$, i.e. smaller galaxies suffered heavier mass loss).

The results of such a simple toy-model for the HMD appear in Fig. 6a. The HMD is extremely well reproduced, down to the lowest metallicities, {\it assuming} $\Phi(M) \propto M^{-2}$, such as those resulting from recent high resolution numerical simulations for Milky Way sized dark haloes (Diemand et al. 2006, Salvadori et al. 2006). Low mass satellites (down to 10$^6$ \ms)  contribute most of the low metallicity stars of the MW halo, whereas the high metallicity stars originate in a couple of  massive satellites with M$\sim$ 10$^8$ \ms \ (Fig. 6a, bottom).

Some properties of the sub-haloes  as a function of their mass appear also in Fig. 6 (right panels).  The outflow rate, in units of the corresponding SFR, is $k(M) = (1-R) (y_{True}/y(M) \ - \ 1)$, where $y_{True}$ is the yield in the solar neighborhood ($y_{true}=0.7 Z_{\odot}$, from the local G-dwarf MD).   
If the MW halo were formed in a a potential well as deep as those of comparable mass galaxies, then  the  large outflow rate required to justify the HMD ($k$=8) is puzzling; on the contrary, the HMD is readily understood if the MW halo is formed from a large number of smaller satellites, each one of them having suffered heavy mass loss according to the simple outflow model.

A more physical, but much less ``transparent'', approach consists in deriving the full merger tree of the MW halo and (by using appropriate receipes for SFR and feedback for the sub-haloes) following the chemical evolution through merging/accretion with Monte-Carlo simulations. Recent studies  (Tumlinson 2006, Salvadori et al. 2006) find good agreement with the observed HMD, but it is hard to find (in vue of the many model parameters) what is (are) the key factor(s) determining the final outcome. In any case, {\it mass loss} (occuring mostly through tidal stripping than through supernova feedback, e.g. Bekki and Chiba 2001) is crucial in shaping the HMD.

\subsection{Implications: abundance dispersion and origin of the r- elements}

The early chaotic history of the MW (according to the hierarchical merging scenario) bears naturally 
to the issue of dispersion in
the abundance ratios of EMP stars. Simple-minded arguments suggest that
dispersion should increase at low metallicities, under the 
explicit assumption
that  low metallicities correspond to such early times that complete 
mixing of SN ejecta with the interstellar medium is impossible
(Argast et al. 2002, Karlsson and Gustafsson 2001). 
In order to  actually see such a dispersion, 
variations in abundance ratios among SN of different masses should be
sufficiently large (say, larger than typical observational statistical 
uncertainties of 0.1 dex).

Several recent studies of EMP stars reveal very small scatter in the abundance ratios of all elements
in the C to Fe-peak mass range
(Carretta et al. 2002, Cayrel et al. 2004, Barklem et al. 2006). This could
mean that i) mixing timescales are (much) shorter than typical chemical
evolution timescales at metallicities down to [Fe/H]$\sim$-3.5 or ii)
variations in abundance ratios of SN of different masses and energies (for the C to Fe-peak mass range)
are sufficiently small, or a combination of (i) and (ii).
Despite our current ignorance of the timescales of mixing
and early chemical evolution, case (i), i.e.  extremely efficient and rapid mixing
of SN ejecta, cannot be the whole truth, since
large variations in abundance ratios are observed in the case of r-elements. A solution to that may be that r-elements originate mostly in a restricted sub-class of CCSN (say 8-10 \ms), which produce a very large r/Fe ratio (e.g. Ishimaru et al. 2003, also these proceedings).

\begin{figure}
\centering
\includegraphics[height=0.8\textwidth,angle=-90]{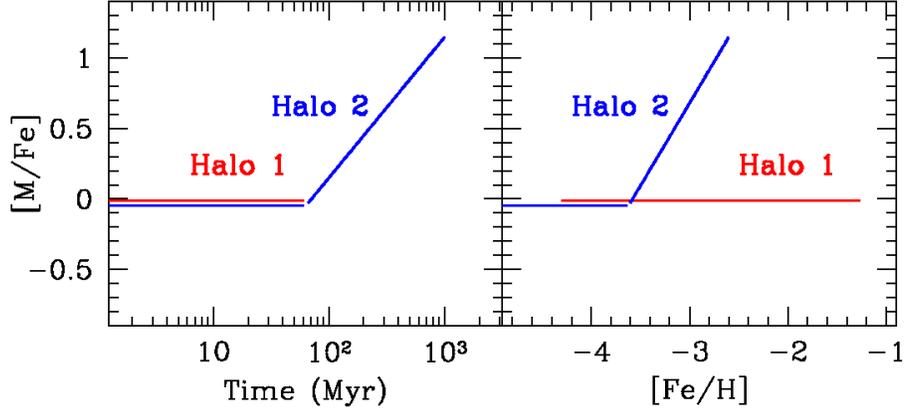}
\caption{Illustration of how differently evolving  sub-haloes may create a dispersion in the abundance ratio [X/Fe], provided element X is produced in two different nucleosynthesis sites, with different evolutionry timescales; this might explain observed dirpersion in r/Fe. {\it Left:} Halo 1 evolves rapidly, so that only the first (short-lived) site of element M operates, and it reaches a high metallicity (say [Fe/H]=-1.2); halo 2 (less massive) evolves on longer timescales, allowing for the second source of element M to increase the M/Fe ratio,  but it reaches a smaller metallicity (say [Fe/H]=-2.6). {\it Right}: When those sub-haloes merge, dispersion as a fuction of metallicity is naturally obtained (starting at low [Fe/H]), even if the ISM in each one of them is completely homogeneized. }
\label{fig:5}       
\end{figure}

The hierarchical merging scenario may complicate even more the already unclear situation of  the r-element production. At present, the
production site of r-nuclei is unknown, with (some class of) CCSN favoured relative to neutron star mergers (NSM). 
The reasons for the ``disfavor'' of NSM are explored and clearly presented in Argast et al. (2004): the combination of
expected low event rates, long coalescence timescales and high individual yields leads to a relatively {\it late injection} of the NSM r-nuclei in the
Galaxy (around [Fe/H]$\sim$--2.5, while Eu is observed already at [Fe/H]$\sim$--3 and the r-component of Ba at even lower metallicities), 
and produces a {\it large scatter of r/Fe ratio relatively late} (at [Fe/H]$\sim$--2, when observations show that such large scatter occurs only at lower metallicities, around [Fe/H]$\sim$--3, Fig. 4 in Argast et al. 2004). 

The conclusions of the Argast et al. (2004) study depend a lot on the adopted assumptions about the NSM properties, as the authors recognize. They also depend on their model of the early Galaxy evolution, in which star formation proceeds stochastically in various places, but {\it at the same average rate everywhere}, i.e. there is in reality one single system evolving. The situation is obviously different in a (presumably more realistic) system composed of many different sub-units, each one evolving at its own rate and affected in a different way by mass loss.



An illustration of such a situation is provided in Fig. 7: Halo 1 (H1, massive) evolves rapidly (100 Myr) and reaches a high metallicity ([Fe/H]$\sim$-1), while halo 2 (H2, low mass) evolves slowly ($\sim$1 Gyr) and reaches a lower final metallicity  
([Fe/H]$\sim$-2.6). Because of short timescales, NSM do not enrich H1 with r-elements, so its r/Fe ratio remains [r/Fe]=const=0. In H2, NSM start increasing the r/Fe ratio only after about 100 Myr (assumed as typical coalescence timescale of NSM), up to [r/Fe]$\sim$1 (before 100 Myr, CCSN produce [r/Fe]=0). Note that efficient and rapid mixing of CCSN and NSM products in the gas of both haloes is always assumed (to ensure no dispersion in abundance ratios {\it within each halo}. When the two haloes merge together, two distinct histories appear in the r/Fe vs Fe/H plane (Fig. 7 right). By considering a large number of intermediate mass haloes, the plane can be naturally filled at [Fe/H]$>$--3. A large dispersion can then be naturally recovered and NSM can be important sources of r-nuclei (along with CCSN), while at the same time dispersion for other elements (exclusively produced in CCSN) remains small, in agreement with observations.

In summary, because of the chaotic early history of the MW (formed by a myriad of smaller sub-haloes with widely differing SF histories), it cannot be excluded at present that both CCSN and NSM have contributed to its enrichment with r-elements.


\end{document}